\documentclass[]{aa}  

\newcommand{\degree}{\ensuremath{^\circ}\xspace}
\newcommand{\Msun}{\ensuremath{\mathrm{M}_{\odot}}\xspace}
\newcommand{\Rsun}
{\ensuremath{\mathrm{R}_{\odot}}\xspace}
\newcommand{\Lsun}
{\ensuremath{\mathrm{L}_{\odot}}\xspace}
\newcommand{\xsource}{XTE J1710--281\xspace}
\usepackage{natbib,twoopt}
\usepackage{multirow}
\usepackage[breaklinks=true]{hyperref}
\usepackage{amsmath}
\usepackage{amssymb}	
\usepackage[T1]{fontenc}
\usepackage{booktabs}
\usepackage{threeparttable}
\usepackage{xspace}
\usepackage{subcaption}
\usepackage[rightcaption]{sidecap}
\usepackage{tabularx}
\usepackage{graphicx}
\usepackage{txfonts}
\bibpunct{(}{)}{;}{a}{}{,} 
\usepackage{siunitx}
\begin{document} 

\title{The puzzling orbital residuals of XTE J1710–281: is  a Jovian planet orbiting around the binary system?}


   \author{R. Iaria\inst{1}, T. Di Salvo\inst{1}, A. Anitra\inst{1}, C. Miceli\inst{1,2}, W. Leone\inst{1,3}, C. Maraventano\inst{1},
   F. Barra\inst{1},  A. Riggio\inst{2,4,5}, A. Sanna\inst{4}, A. Manca\inst{4},  and  L. Burderi\inst{2,4}}

 \institute{Dipartimento di Fisica e Chimica - Emilio Segrè,
 Universit\`a di Palermo, via Archirafi 36 - 90123 Palermo, Italy
 \and
             INAF/IASF Palermo, via Ugo La Malfa 153, I-90146 Palermo, Italy
  \and
 Department of Physics, University of Trento, Via Sommarive 14, 38122 Povo (TN), Italy 
  \and
         Dipartimento di Fisica, Universit\`a degli Studi di Cagliari, SP
Monserrato-Sestu, KM 0.7, Monserrato, 09042 Italy 
\and
INFN, Sezione di Cagliari, Cittadella Universitaria, 09042 Monserrato, CA, Italy
  }

  \abstract
   {}
   {\xsource   is a transient  eclipsing binary system with a period close to 3.28 hours, hosting a neutron star. The average eclipse duration is 420 seconds, and eclipse arrival times reported in the literature span from 1999 to 2017. A previous analysis of the eclipse arrival times using the eclipse timing technique revealed a complex pattern of delays, indicating the presence of three orbital glitches. These glitches correspond to sudden variations in the orbital period, allowing for the identification of four distinct epochs
   during which the orbital period derivative was estimated to be 
   $-1.8 \times 10^{-12}$ s~s$^{-1}$, $0.07 \times 10^{-12}$ s~s$^{-1}$,  $-1.8 \times 10^{-12}$ s~s$^{-1}$ and  $0.09 \times 10^{-12}$ s~s$^{-1}$, respectively. 
   }
   {We have re-analyzed the 78 eclipse arrival times spanning 18 years
   utilizing the eclipse timing technique to derive the corresponding delays as a function of time.}
   {We find that the observed delays align well with a fitting model employing an eccentric sine function characterized by an amplitude of $6.1 \pm 0.5$ s, eccentricity of $0.38 \pm 0.17$, and a period of $17.1 \pm 1.5$ years. Additionally, we  identified the orbital period as 3.28106345(13) hours, with a reference epoch of $T_0=54112.83200(2)$ Modified Julian Date (MJD). We obtained an upper limit of the orbital period derivative of $3.6 \times 10^{-13}$ s~s$^{-1}$.}
  {From the average value of the eclipse duration, we estimate that the companion star has a mass of 0.22~\Msun for a neutron star mass of 1.4~\Msun, and the inclination of the source is $78.1^{+1.5}_{-1.2}$ degrees. The companion star is in thermal equilibrium. The orbital period derivative is consistent with a conservative mass transfer scenario, where the angular momentum loss due to magnetic braking dominates over gravitational radiation angular momentum loss  if the former is present. The eccentric modulation can be explained by a third body with a mass of 2.7 Jovian masses, orbiting with a revolution period close to 17 years and an eccentricity of 0.38.}

  \authorrunning{R. Iaria et al.}

  \titlerunning{Is a Jovian planet orbiting around XTE J1710–281?}
  
  

   \maketitle
%

\section{Introduction}
\xsource, discovered in 1998 by the \textit{Rossi X-ray Timing Explorer} (RXTE), is a transient low-mass X-ray Binary (LMXB) likely associated with the \textit{ROSAT} source 1RXS J171012.3-280754 \citep{Mark_98}. This source exhibits high variability, and numerous Type-I bursts have been documented \citep{Mark_01,Galloway_08},  indicating that the compact object is a neutron star. 
\citealt{Galloway_08} estimated, based on Type-I bursts, that the distance to the source is 12~kpc assuming accreting matter with cosmic abundances, or 16~kpc for accreting matter with helium only.

\xsource has an orbital period of 3.28 hours \citep{Mark_01}, inferred from the analysis of the total eclipses observed in its light curve, which also  shows regular dipping phenomena, potentially attributed to occultation from the outer regions of the accretion disc. The presence of dips and total eclipses suggests that the system has an inclination angle between 75\degree and 80\degree \citep{Frank_87}. The dips in \xsource have been studied by \cite{younes_09}; analyzing \textit{XMM-Newton} data of the source they observed that the hydrogen column density is $\sim 4 \times 10^{21}$~cm$^{-2}$ during the persistent emission, from $ 4 \times 10^{21}$~cm$^{-2}$ to $ 8 \times 10^{23}$~cm$^{-2}$ during shallow dips and $\sim 1.4 \times 10^{23}$~cm$^{-2}$ during deep dips. 
\begin{figure*}[]
    \centering
    \begin{subfigure}{.5\textwidth}
        \centering
        \includegraphics[width=.92\linewidth]{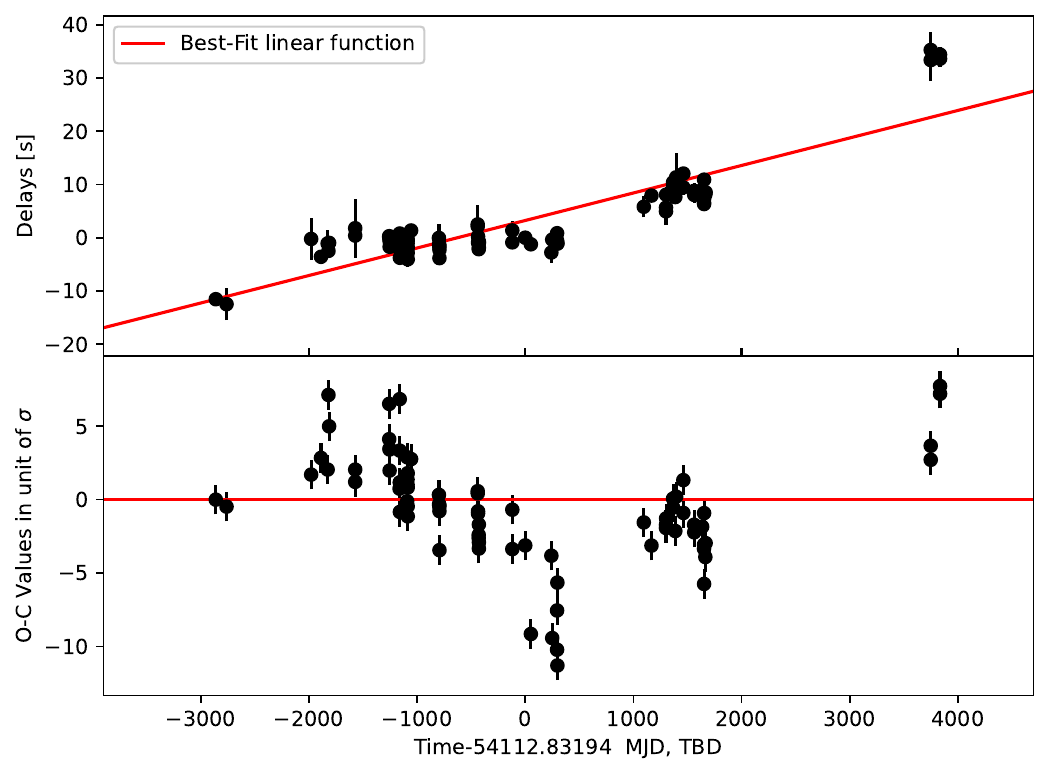}
        
            \caption{}
      \label{fig:figure1a}
    \end{subfigure}%
    \begin{subfigure}{.5\textwidth}
        \centering
        \includegraphics[width=.905\linewidth]{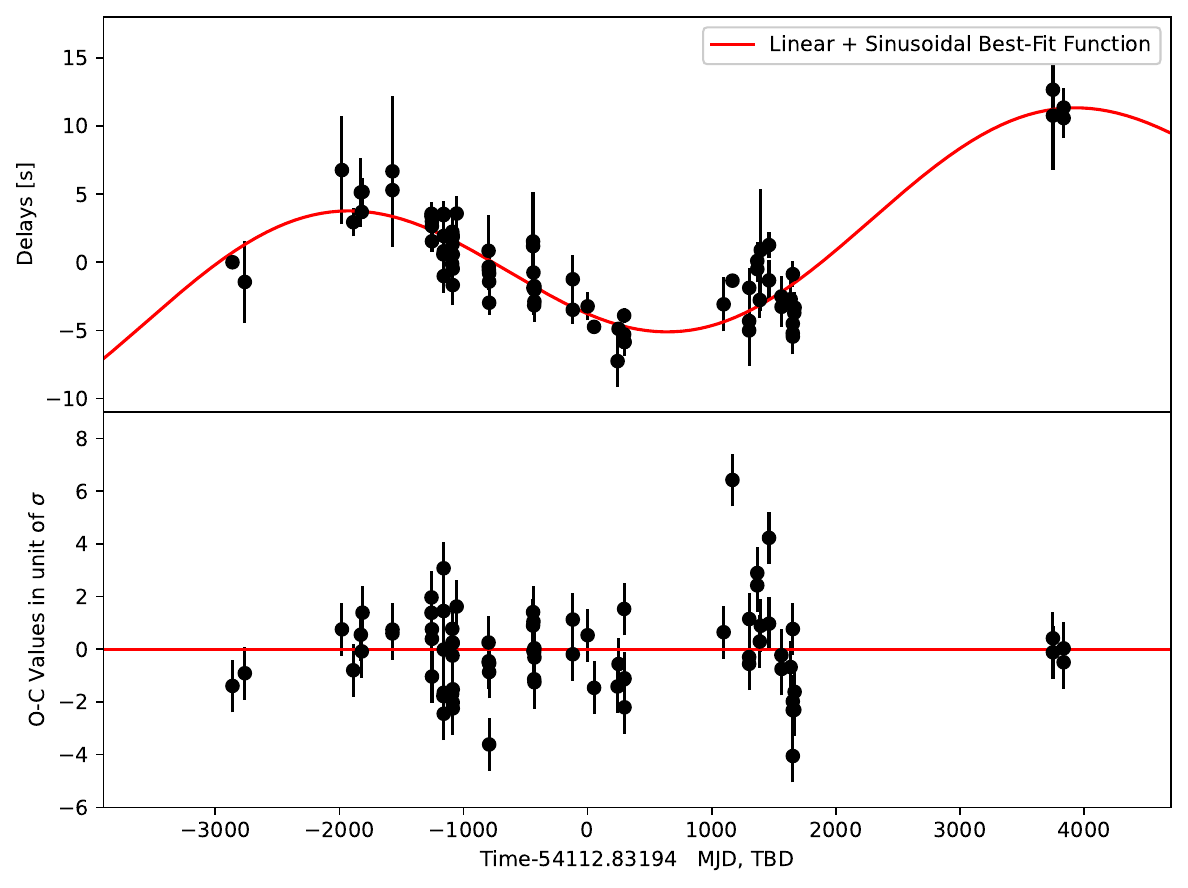}
             \caption{}
      \label{fig:figure1b}
    \end{subfigure}
    \caption{(a) Mid-eclipse delays in units of seconds   as a function of time obtained adopting as reference times $T_0=54112.83194$~MJD and as orbital period the  value of $P_0= 3.2810632$~h (top panel). The best-fit linear function (red color) is described in the text;  O-C values are expressed in units of $\sigma$ (bottom panel). (b) Mid-eclipse delays in units of seconds  obtained adopting as reference times $T_0=54112.831977$~MJD and orbital period the arbitrary value of $P_0= 3.28106339$~h (top panel). The best-fit curve  is described by Eq. \ref{eq:sine}.  O-C values are expressed in units of $\sigma$ (bottom panel).}
    \label{fig:linear}
\end{figure*}
 
Analyzing \textit{Chandra} and \textit{Suzaku} data, \cite{Raman_18} observed a broad emission line at 0.72~keV, associated with  partially ionized iron, and an absorption line in the Fe-K region of the spectrum at 6.6~keV, interpreted as a blend of \ion{Fe}{xix} to \ion{Fe}{xxv} transitions. 

An accurate study of the eclipses of \xsource was performed by \cite{Chetana_11} and \cite{chetana_22} which derived 78 mid-eclipse arrival times analyzing RXTE,  \textit{Chandra}, \textit{Suzaku}, \textit{XMM-Newton} and \textit{AstroSat} observations of the source spanning 18 years. The authors estimated the duration of the eclipse to be 420~s on average and, studying the delays with respect to the orbital cycles, they reported the presence of three orbital glitches 
that current theoretical models cannot explain.

X-ray binaries undergo evolutionary processes driven by various physical mechanisms. One major factor influencing their evolution is mass transfer. This process leads to the re-distribution of angular momentum, impacting the orbital dynamics \citep{heuvel_94}.
Moreover, a non-conservative mass transfer can occur as a consequence of different processes, such as 
radiative evaporation of the secondary star or the expulsion of matter in the form of accretion disc winds, outflows, or jets \citep{ruderman_89,Brookshow_93,Ponti_2012}. These mechanisms alter the overall mass content and momentum distribution within the system.

Furthermore, angular momentum lost in the system occurs through two  mechanisms: gravitational wave radiation and magnetic braking of the tidally locked companion star \citep{rappaport_83, Applegate_92, Applegate_94, verbunt_93}."
These processes directly impact the orbital parameters, contributing to the dynamic evolution of X-ray binaries over time. Consequently, the orbital separation in X-ray binaries may either increase or decrease.

In this study, we exploit the mid-eclipse arrival times obtained from \cite{Chetana_11} and \cite{chetana_22} to propose a different  interpretation for the orbital residuals of \xsource, suggesting that the eclipse arrival times are affected by the presence of a third body orbiting around the binary system with a mass of 2.7 Jupiter masses and an orbital eccentricity of 0.38.

\section{Data analysis}
\begin{figure*}[!htbp]
   \centering
   \begin{subfigure}{.5\textwidth}
      \centering
      \includegraphics[width=.92\linewidth]{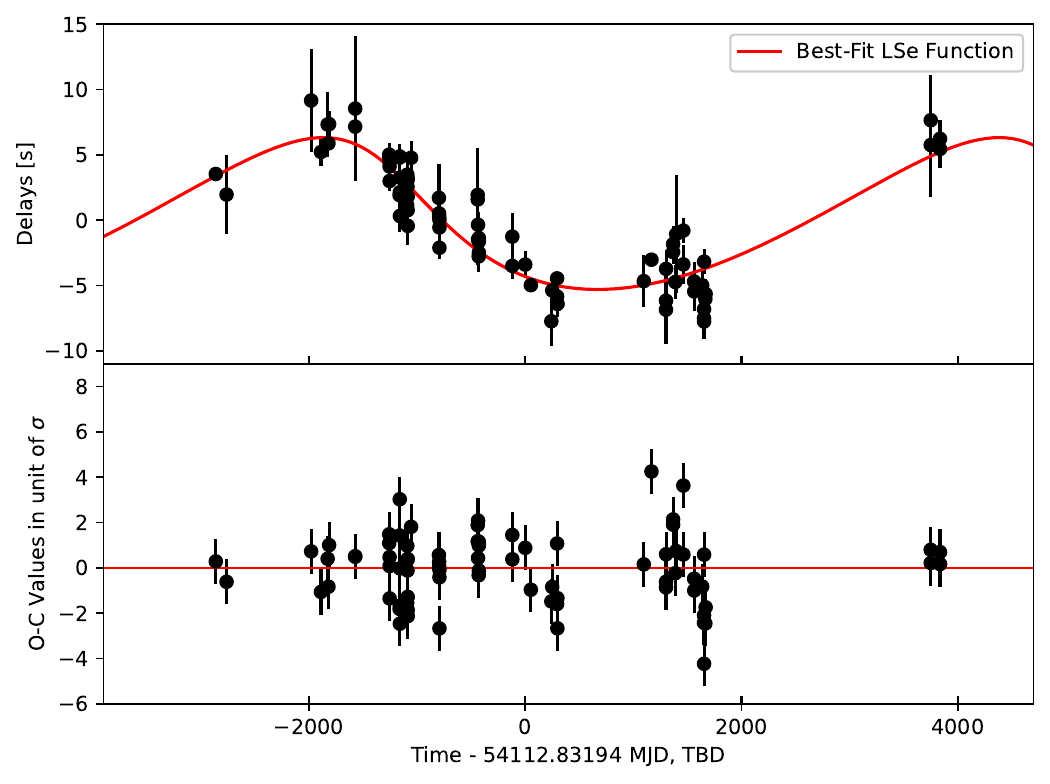}
      \caption{}
      \label{fig:subelliptical}
   \end{subfigure}%
   \begin{subfigure}{.5\textwidth}
      \centering
      \includegraphics[width=0.905\linewidth]{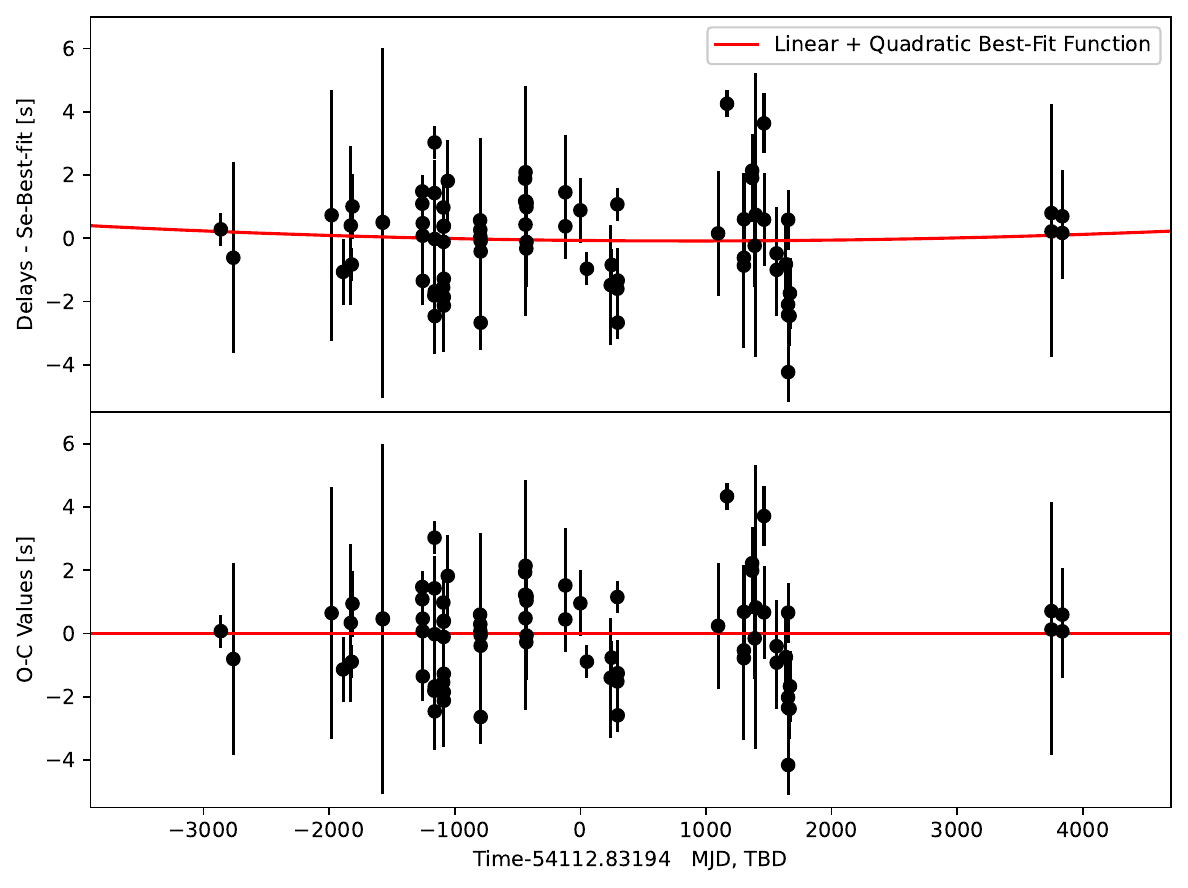}
      \caption{}
      \label{fig:subquad}
   \end{subfigure}
   \caption{(a) Mid-eclipse delays obtained adopting as reference times $T_0=54112.831979$~MJD and as orbital period the  value of $P_0= 3.28106345$~h (top panel). The best-fit curve (red color) is described in the text. O-C values in units of $\sigma$ (bottom panel). (b) Mid-eclipse delays are obtained by subtracting the eccentric sinusoidal modulation (top panel). The best-fit curve is described by Eq. \ref{eq:quad}.  Corresponding O-C values in seconds (bottom panel).}
   \label{fig:elliptical}
\end{figure*}
\begin{table} 
\setlength{\tabcolsep}{3.5pt}
        \centering
        \caption{Best-fit parameters.} 
                \label{tab:fit}
        \begin{tabular}{l | c |c}
        \toprule
          Parameters   & LS function
                       &   LSe function\\    
        \midrule
         $q$ (s)&   
          0.16(0.59) &  1.9(1.4)\\

          $m$ (s/days) &   
          0.00129(34) & [0]  \\

         $A$ (s)&   
          6.2(7) & 6.1(5) \\

          $P_m$ (days)&   
        5851(576) & 6249(564) \\

         $t_0$ (days)&   
        2285(257)   & -- \\  

         $T_p$ (days)&   
        -- & -1261(322)   \\  

          $e$ &   
        -- & 0.38(17)   \\  
          $\omega$   (rad)&   
        -- & 2.5(4)   \\  

& & \\
        
          $T_0$ (MJD,TBD)&   
        54112.831979(7)    & 54112.83200(2) \\  
          $P_0$ (h)&   
        3.28106345(13)    &  3.28106345 \\  
        & & \\
        
            $\chi^2$/d.o.f.  
             & 253/73  &      
            173/72   \\ 
 \bottomrule
\end{tabular}
    \begin{tablenotes}
\item[]    Errors are given at a 68\% confidence level. The value in square brackets is kept fixed during the fit. 
    \end{tablenotes}
    \end{table}

We used the 78 mid-eclipse arrivals times shown by \cite{Chetana_11} and \cite{chetana_22}. We started assuming as reference orbital period $P_0=3.2810632$~h and as reference time $T_0=54112.83194$~MJD corresponding to the mid-eclipse time observed by the RXTE observatory during the observation ID. 91018-01-02-00 \citep[see Table 1 in][]{Chetana_11}. We inferred the delays corresponding to each eclipse time by taking the fractional part of $(T_{ecl}-T_0)/P_0$ and multiplying it by the value of  $P_0$ expressed in seconds. We show the eclipse times and the corresponding delays in the top panel of Fig. \ref{fig:figure1a}.

 To correct the values of $P_0$ and $T_0$, we fitted the delays with a linear function ${y=m\;(t-T_s)+q}$. The term ${m=\Delta P/P}$ gives the correction to $P_0$, ${q=\Delta T_0}$ is the correction to the reference time $T_0$ and, finally, $T_s$ is fixed to 54112.83194~MJD and corresponds to a temporal shift applied to the eclipse times.

From the fit we obtained  a $\chi^2$(d.o.f.) of  1298(76),  $m=0.0052(3)$ s/days and $q=3.2(5)$ s. The errors are reported at a 68\% ($1\sigma$) confidence level. We show the linear best-fit in the top panel of  Fig. \ref{fig:figure1a}. The observed minus calculated (O-C) values in units of $\sigma$ displayed in the bottom panel of Fig. \ref{fig:figure1a} show that the correction is not sufficient to describe the temporal evolution of the source as the delays deviate from the best-fit model up to 10~$\sigma$. However, a sinusoidal modulation seems to be present in the residuals.  

We implemented the corrections from the linear fit by obtaining $P_0=3.281063396(11)$~h and $T_0=54112.831977(6)$~MJD and recalculated the delays corresponding to the mid-eclipse times accordingly. We plot the delays versus time in the top panel of  Fig.
\ref{fig:figure1b}. To fit the delays, we adopted a linear plus sinusoidal function (hereafter LS function) defined as:
\begin{equation}
\label{eq:sine}
  y= q + m \;(t-T_s)+A \sin\left[2\pi \frac{[(t-T_s)-t_0]}{P_m}\right] ,
\end{equation} 
where the sinusoidal term takes into account the modulation observed in the O-C values (bottom panel of Fig. \ref{fig:figure1a}). 
By fitting the data, we found a $\chi^2$(d.o.f.)  of  253(73) that translates to a F-test probability of chance improvement of $7.4 \times 10^{-26}$ compared to the linear model.  The associated errors of the best-fit parameters were scaled by the factor $\sqrt{\chi^2_{red}}$
to take into account a value of the $\chi^2_{red}$ of the best-fit model larger than one. 
We found the following best-fit values: $q=0.16(59)$~s, $m=0.00129(34)$~s/days, $A=6.2(7)$~s, $P_m=5851(576)$~days (i.e., $16.0\pm 1.6$~years) and $t_0=2285(257)$~days.  By correcting $P_0$ and $T_0$ with the best-fit values of $m$ and $q$, we obtain $T_0=54112.831979(7)$~MJD and $P_0=3.28106345(13)$~h. We show the best-fit function 
in red in the top panel of Fig. \ref{fig:figure1b} and the corresponding O-C values in units of sigma in the bottom panel. 
We report the best-fit values of the parameters in the second column of \autoref{tab:fit}.

The ephemeris of the source obtained using the LS function is:
\begin{equation*}
\begin{split} 
 { \rm {T_{ecl}= MJD(TDB)\; 54112.831979(7) + \frac{3.28106345(13)}{24}N +}}\\
{\rm \frac{6.2(7)}{86400}\sin\left[2\pi \frac{(t-54112.83194)-2285(257)}{5851(576)}  \right],  }
\end{split}
\end{equation*} 
where $N$ indicates the orbital cycle and $t$ is expressed in MJD.

We recalculated the delays using the best-fit LS values of $T_0$ and $P_0$. We then fitted the delays with the LS function, ensuring that the parameters $q$ and $m$ had best-fit values compatible with zero. The other fit parameters yield the same best-fit values as reported in the second column of \autoref{tab:fit}. 
In light of the significantly high $\chi^2$/d.o.f. value, attributable to residuals that still exhibit clear deviations from the LS model, we considered the possible scenario in which the sinusoidal modulation may be associated with the presence of a third body orbiting around the binary system with eccentricity non-null in a hierarchical triple system. 

The determination of the orbital eccentricity was achieved through the application of a numerical solution to Kepler's equation. The utilized technique involved modeling the delay modulation  with an elliptical orbit. The eccentricity, $e$, was extracted by employing an iterative approach to solve Kepler's equation. The iterative solution, known as the Newton-Raphson method, facilitated the computation of the eccentric anomaly $E=M-e\sin{E}$ (where $M$ is the mean anomaly), providing a quantitative measure of  $e$. 
$M$ is defined as $\pi [(t-T_0-T_s)-T_p]/P_m$, where $T_p$ is the periastron passage time and $P_m$ the period of the the delay modulation.

We called LSe the function adopted to fit the delays, which includes a constant term ${q}$, a linear term ${m}$, and an eccentric sinusoidal term. We fitted the data, obtaining a $\chi^2$(d.o.f.) of 171.8(71) and a value of the parameter ${m}$ of $3.3(49) \times 10^{-2}$~s/days that is compatible with zero as we expect. To avoid issues related to parameter degeneracy, we fixed the value of ${m}$ to zero, assessing later its influence on the $\chi^2$ value.

By fitting the delays with the LSe model we found a $\chi^2$(d.o.f.) of 173(72) and $\Delta \chi^2$
of 80 with respect to the best-fit obtained using the LS function; the f-test probability of chance improvement is $1.85 \times 10^{-7}$, corresponding to a significance of 5.2 $\sigma$, suggests that a non-null eccentricity is required 
with high statistical significance.
The associated errors of the best-fit parameters were scaled by the factor $\sqrt{\chi^2_{red}}$
to take into account a value of $\chi^2_{red}$ of the best-fit model larger than one. We obtained 
$q=1.9(1.4)$~s,  $A=6.1(5)$~s, $P_m=6249(564)$~days corresponding to $17.1\pm 1.5$ years,
$T_p=-1261(322)$~days, $e=0.38(17)$ and $\omega=2.5(4)$~rad, where $\omega$ is the argument of periastron. The delays, the best-fit function, and the corresponding O-C values in units of $\sigma$ are shown in Fig. \ref{fig:subelliptical} (top and bottom panel, respectively). The best-fit values of the parameters are shown in the third column of \autoref{tab:fit}. 
Setting the parameter ${m}$ equal to zero has little influence on the fit; the f-test probability of chance improvement, leaving the parameter free to vary, is of the order of 0.5.

 \begin{figure*}
  \includegraphics[width=5cm]{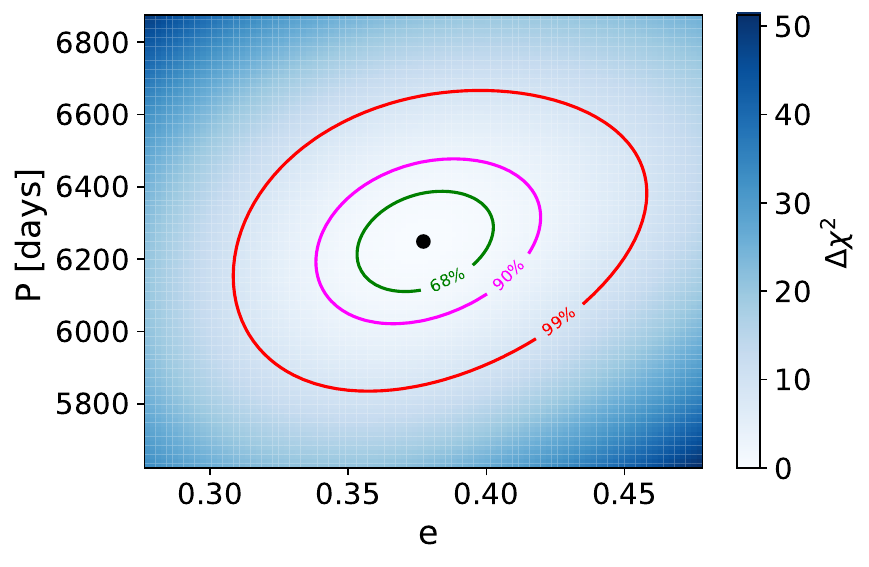}
  \includegraphics[width=5cm]{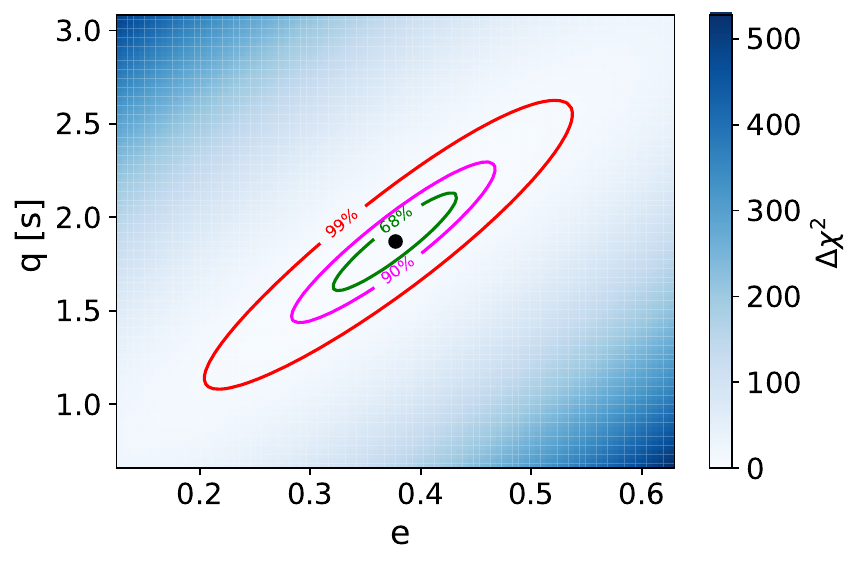}   
    \includegraphics[width=5cm]{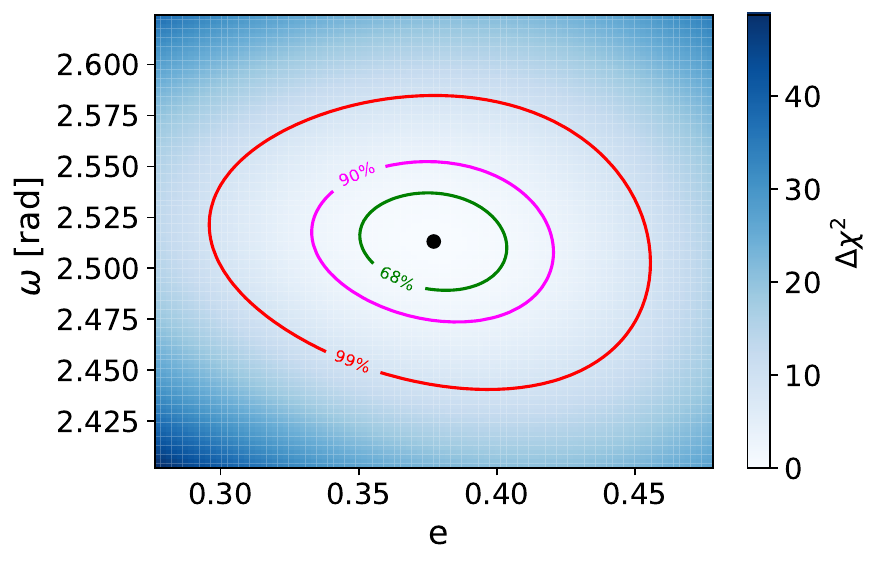}\\
                \sidecaption  
      \includegraphics[width=5cm]{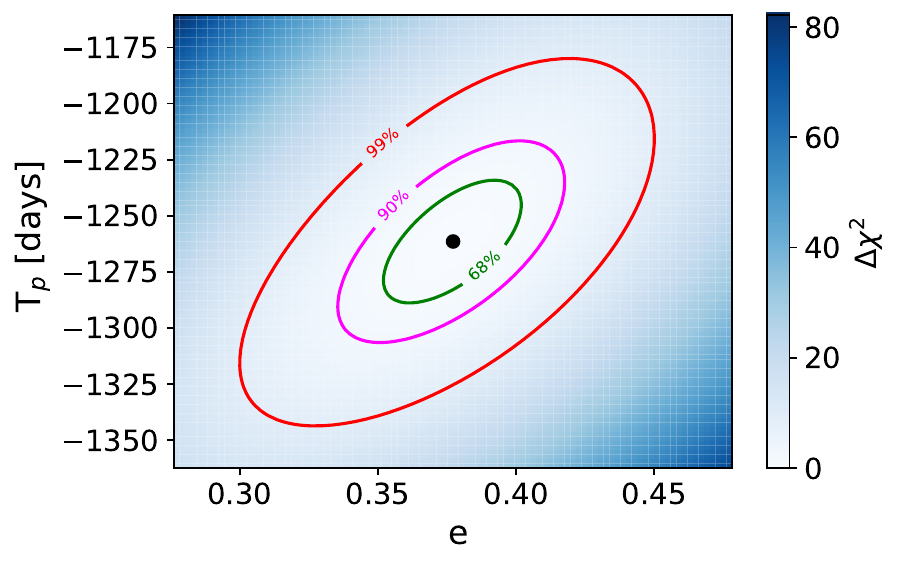}   
            \includegraphics[width=5cm]{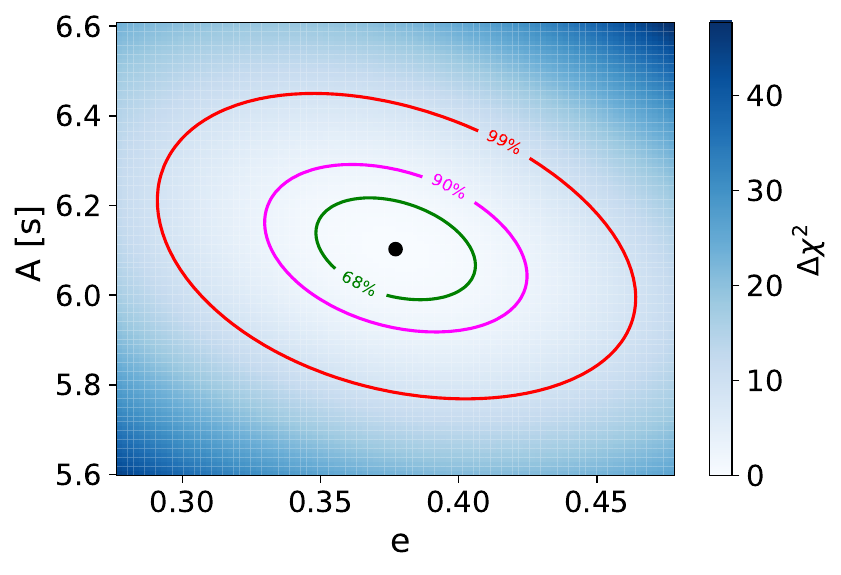}  
  \caption{Correlation of eccentricity (${e}$) with other parameters of the LSe function. The correlations are modest, with the most notable observed between parameters ${q}$ and ${e}$. The eccentricity is constrained to be within 0.3 and 0.5 at a 90\% confidence level. Contours in green, magenta, and red represent the 68\%, 90\%, and 99\% confidence levels.}
     \label{fig:correlation}
\end{figure*}

We also explored the correlation of the eccentricity with other parameters but with the parameter ${m}$ fixed to zero. This correlation is illustrated in Fig. \ref{fig:correlation}, where contour plots for 68\% (green), 90\% (magenta), and 99\% (red) confidence levels are shown across all six panels. Notably, the strongest correlation is observed between eccentricity and parameter ${q}$, as well as between the amplitude ${A}$ of the sinusoidal function and eccentricity ${e}$, where ${e}$ varies within the range of 0.25 to 0.55. The inclusion of eclipse arrival times distributed over a broader temporal baseline will serve to minimize these correlations, enhancing the reliability of our findings.

To obtain a constraint on the derivative of the orbital period $\dot{P}$, we subtracted the best-fit eccentric sinusoidal function from the delays. We then fitted the modified delays using the quadratic function
\begin{equation}
\label{eq:quad}
  y= q + m\;(t-T_s)+ c\;(t-T_s)^2 ,
\end{equation} 
where  $c=\dot{P}/(2P_0)$ in s/days$^2$. 
We show the data and the best-fit function in the top panel of Fig. \ref{fig:subquad}. 
We obtained the following best-fit values: $q=-0.07 \pm 0.33$~s, $m=(-0.4\pm 2.0) \times 10^{-4}$~s/days and $c=(2\pm9) \times 10^{-8} $~s/days$^2$. Adopting the best-fit value of $c$ we found that $\dot{P}=(0.7\pm 2.9)\times 10^{-13}$~s~s$^{-1}$. We show the O-C residuals with respect to this model in the bottom panel of Fig. \ref{fig:subquad}. 

\section{Discussion}
 
\subsection{The jittered behavior in the eclipse arrival times}

In our analysis, we investigated the total eclipse arrival times derived by \cite{Chetana_11} and \cite{chetana_22}, who utilized a step-and-ramp function to model the eclipse shape in the light curves of XTE J1710-281. 
However, it is worth noting that the eclipse shape may vary from one eclipse to another, leading to changes in its ingress/egress and duration times, or solely its ingress/egress time.
 The ingress, egress, and eclipse durations show a jittered behavior of the order of 15 s  in EXO 0748-676  \citep{Wolff_02}, close to 5 s  in MXB 1659-298 \citep{iaria_18} and up to 10 s  in AX J1745.6-2901 \citep{Ponti_17}.  We show in the bottom panel of Fig. \ref{fig:subquad} that the jittered behavior is close to 5 s  for XTE J1710-281. 

\cite{Wolff_07} proposed that the magnetic activity of the companion star can generate extended coronal loops above the companion star's photosphere,  accounting for the observed jitters.
Furthermore, \cite{Ponti_17}  presented a different scenario for AX J1745.6-2901, where jitters were observed in the ingress and egress, while the eclipse duration remained relatively constant.
 They hypothesized that matter ejected from the accretion disc can interact with the companion star's atmosphere, causing displacement and hence delays in the ingress and egress times.

Both scenarios suggest that the presence of magnetic activity and/or the atmosphere of the companion star may slightly alter the estimates of the eclipse arrival times, contributing to the reduced chi-square, which may result  larger than one, as obtained for our best-fit model.

\subsection{Constraints on the companion star mass}
\xsource shows total eclipse and dips, therefore its inclination angle must be between 75\degree and 80\degree  \citep[see][and references therein]{younes_09}. Moreover, the duration of the eclipse $\Delta T_{ecl}$ is on average about 420 seconds \citep{Chetana_11}. Using these observational results, we can provide an estimate of the mass ratio $q=M_2/M_1$, where $M_1$ and $M_2$ are the masses of the neutron star (NS) and the companion star (CS).  
\begin{figure}[!htbp]
   \centering
    \includegraphics[scale=.6]{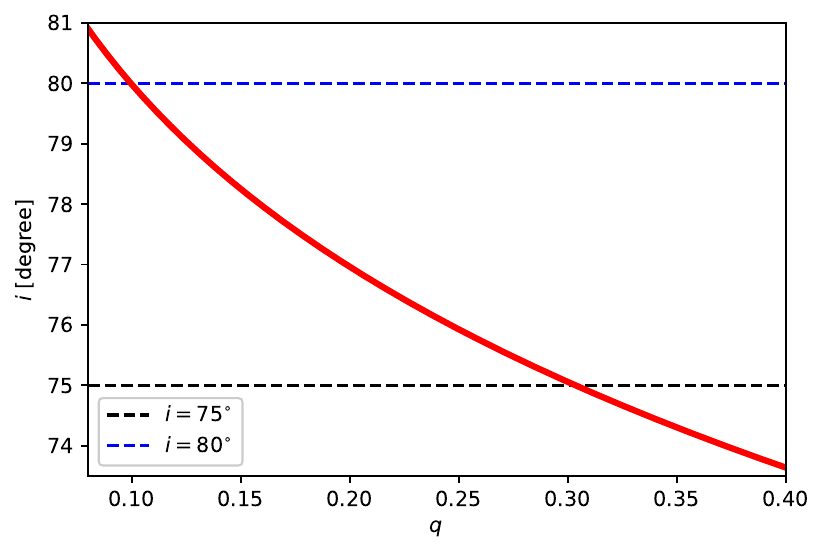}
    \caption{Variation of the inclination angle with the mass ratio $q$ (red curve) as described by Eq. \ref{eq:theta}.}
   \label{fig:i_vs_q}
\end{figure}

Knowing that the eclipse duration is $\Delta Tecl =420$~s, we can estimate the size of the occulted region $x$ \citep[see Figure 5 in][]{iaria_18} by using
the expression:
\begin{equation}
\label{eq:x}
\frac{2 \pi}{P}a = \frac{2x}{\Delta T_{ecl}},
\end{equation}    
where $P$ and $a$ are the orbital period and the orbital separation of the system. The value of $x$ depends on $a$, which is unknown.  The angle $\theta$, complementary to the inclination angle $i$ of the binary system, can be estimated using the following relationship:
\begin{equation}
\label{eq:theta}
\tan\theta = \left[ \frac{R_2^2-x^2}{a^2-(R_2^2-x^2)}\right]^{1/2}, 
\end{equation}  
where $R_2$ is the CS radius. In light of the system accreting via the inner Lagrangian point, the CS fills its lobe, resulting in the radius of the CS coinciding with the Roche lobe radius of the CS, which is given by \citep{Eggleton_83} 
\begin{equation}
\label{eq:rl2}
R_{L2}=a \frac{0.49 q^{2/3}}{0.6q^{2/3}+\ln(1+q^{1/3})}. 
\end{equation} 
 By substituting Eqs. \ref{eq:x} and \ref{eq:rl2} into Eq. \ref{eq:theta}, the angle $\theta$ depends solely on $P$, which is the known orbital period, on $\Delta T_{ecl}$, which is 420 s, and on $q$, our unknown variable. We show the dependence of the inclination angle on $q$ in Fig. \ref{fig:i_vs_q}. 

Knowing that the inclination angle of the system ranges between 75\degree and 80\degree, we deduce that the mass ratio $q$ falls within the range of 0.1 to 0.3. Assuming a NS mass of 1.4~M$_{\odot}$, the CS mass ranges between 0.14~M$_{\odot}$ and 0.43~M$_{\odot}$ for an inclination angle of 80\degree and 75\degree, respectively.

We can estimate the CS mass with further precision using the mass-radius relationship for stars in thermal equilibrium derived from the study of cataclysmic variables in \cite{Knigge_11}. Below, we adopt the mass-radius relationship for binary systems with an orbital period greater than 3.18 hours (the orbital period of \xsource is close to 3.28 h). The relationship is 
\begin{equation}
\label{eq:kingge}
R_2 = 0.293\pm 0.010\;\left(\frac{M_2}{M_{\rm conv}} \right)^{0.69\pm0.03}  {\rm R_{\odot}}, 
\end{equation}  
where M$_{\rm conv}$ has a value of $0.20 \pm 0.02$~\Msun and represents the mass of the convective region of the CS. The latter fills its Roche Lobe $R_{L2}$, so we can assume  that $R_2=R_{L2}$. 
Since we estimated that $q$ is between 0.1 and 0.3, we can use the expression for the Roche lobe radius proposed by \cite{Pac_71}. Combining the expression for the Roche lobe radius with Kepler's third law, we find that:
\begin{equation}
\label{eq:combined}
R_{L2} = 0.233 m_2^{1/3} P_h^{2/3}  {\rm R_{\odot}}, 
\end{equation}  
where $m_2$ is the CS mass in units of solar masses, and $P_h$ is the orbital period in hours. 
Combining Eqs. \ref{eq:kingge} and \ref{eq:combined}, we find M$_2=0.22\pm0.07$~\Msun.

Assuming a NS mass of 1.4~\Msun, we infer that  $q=0.16\pm0.05$ corresponds to an inclination angle of
$i = 78.1^{+1.5}_{-1.2}$ degrees. 
For a NS mass of 2~\Msun we obtain $q=0.11\pm 0.03$ corresponding to an inclination angle of $i = 79.6^{+1.6}_{-1.2}$ degrees.

 Using Kepler's third law, we estimate that the orbital separation of \xsource is $a = 9.1 \times 10^{10}$~cm for an inclination angle of 78$^{\circ}$, $M_1=1.4$~\Msun, and $M_2=0.22$~\Msun. 

\subsection{Self-consistence test
for the thermal equilibrium of the CS}

We can use the bolometric X-ray luminosity of the source to estimate whether the CS is in thermal equilibrium. To do this, we impose that the Kelvin-Helmholtz time-scale $\tau_{KH}$ (corresponding to the characteristic time that a star takes to reach thermal equilibrium) is equal or less than the mass transfer time-scale $\tau_{\dot{M}}$. The Kelvin-Helmholtz time-scale is given by the expression
\begin{equation}
\label{eq:KH}
\tau_{KH}=3.1\times 10^7\left( \frac{M_2}{\Msun}\right)^2\left(\frac{\Rsun}{R_2}\right)\left(\frac{\Lsun}{L_2}\right)~{\; \rm year},
\end{equation}
\citep{verbunt_93}. Adopting the mass-luminosity relation for M-type stars proposed by \cite{Neece_84} and given by the expression $L_2/\Lsun=0.231 (M_2/\Msun)^{2.61}$, along with the mass-radius relation from Eq. \ref{eq:kingge}, we obtain $\tau_{KH} \simeq 1.5 \times 10^8 \; m_2^{-1.3}$ year. The mass-transfer time-scale is given by 
\begin{equation}
\label{eq:tmdot}
\tau_{\dot{M}}=\frac{M_2}{\dot{M}_2}=\frac{GM_1M_2}{L_XR_{NS}},  
\end{equation}
where $\dot{M}_2$ is the mass transfer rate, $G$ is the gravitational constant, $L_X$ is the X-ray bolometric luminosity and $R_{NS}$ is the NS radius. In the last equality of Eq. \ref{eq:tmdot}, we imposed that the mass transfer rate is equal to the mass accretion rate in the scenario of conservative mass transfer. This point is further discussed later on. 
We find that $\tau_{\dot{M}} \simeq 3.7 \times 10^{53} \;m_2 L_X^{-1}$ s for a NS radius of 10 km and a NS mass of 1.4~\Msun.

By imposing that $\tau_{KH} \leq \tau_{\dot{M}}$, we obtain $m_2 \geq (1.28 \times 10^{-38} \;L_X)^{1/2.3}$. The unabsorbed luminosity in the 0.2-10 keV  energy band of \xsource, during the persistent emission, was estimated to be  $L_X \simeq 2.4 \times 10^{36}$ erg/s for a distance to the source of 16~kpc  \citep{younes_09}. For this luminosity value, we determine that the CS is in thermal equilibrium when $m_2 \geq 0.22$, which is the CS mass determined by our calculations.

\subsection{The orbital period derivative of \xsource}

Drawing upon theoretical frameworks, the mass-transfer rate $\dot{M}_2$ within the context of long-term orbital evolution  
can be expressed as 
\begin{equation}
\label{eq:pdot}
\dot{m}_{-8}=35\; (3n-1)^{-1}m_2\frac{\dot{P}_{-10}}{P_{5h}},
\end{equation}  
where $\dot{m}_{-8}$ is the mass transfer rate in units of $10^{-8}$~\Msun/year, $n$ is the mass-radius index of the CS, $\dot{P}_{-10}$ is the orbital period derivative in units of $10^{-10}$~s/s and $P_{5h}$ is the orbital period in units of 5 hour \cite[see][and references therein]{Burderi_10}.

Assuming a conservative mass-transfer $\dot{M_1}=-\dot{M_2}$ and considering a source luminosity of $2 \times 10^{36}$~erg/s,
in the hypothesis that the observed luminosity is a good tracer of the mass accretion rate $\dot{m}$, we find $\dot{m}=1.7 \times 10^{-10}$~\Msun/year for a NS mass of 1.4 \Msun and a NS radius of 10 km. Consequently, the mass transfer rate is $\dot{M_2} = -1.7 \times 10^{-10}$~\Msun/year.
Using $n=0.69$ (adopted in Eq. \ref{eq:kingge}),  $m_2=0.22$ and an orbital period of 3.28~h, we find $\dot{P} = -1.5 \times 10^{-13}$~s~s$^{-1}$. This value is consistent within  1$\sigma$ with what we obtained from our analysis, that is  $\dot{P} = (0.7\pm 2.9) \times 10^{-13}$~s~s$^{-1}$; therefore, we should expect that the binary system is undergoing a contraction of the orbit. 

The orbital period change due to the loss of angular momentum can be associated with the emission of gravitational waves (GR) and magnetic braking (MB). Below, we explore both cases assuming a conservative mass-transfer scenario.  The orbital period changes due to GR is given by the relationship 
\begin{equation}
\label{eq:pd_grav}
\begin{split} 
{\dot{P}_{\rm grav} = -1.4\times 10^{-12} m_1 m_2 m_T^{-1/3} P_{2h}^{-5/3} [(n-1/3) \times} \\
{(n+5/3-2q)]\;\; {\rm s~s^{-1}}},  
\end{split} 
\end{equation} 
where $m_1$ is the NS mass in units of \Msun, $m_T$ is $m_1+m_2$ in units of \Msun, $P_{2h}$ the orbital period in units of two hours and $q$ is the mass ratio of the binary system \cite[see][and references therein]{disalvo_08}. Using the values of $m_1$, $m_2$, $P$ and $n$ shown above, we estimate $\dot{P}_{\rm grav} \simeq  -2.8 \times 10^{-14}$~s/s. 

The estimated mass of the CS is less than 0.3~\Msun. \cite{rappaport_83} proposed that the magnetic braking mechanism might not be active, attributing this to the star being fully convective. The lack of a radiative zone implies 
losing its magnetic field dipolar structure. If this mechanism is inhibited, the only contribution to the orbital period derivative is related to $\dot{P}_{grav}$. However, both \cite{Chen_17} and \cite{Tailo_18} indicated that the magnetic braking mechanism could remain active even in stars with masses below 0.3~\Msun. In such a scenario, we expect the contribution of magnetic braking to the orbital period variation to be significant. 
 
The contribution to the orbital period variation related to magnetic braking is given by the relation:
\begin{equation}
\label{eq:pd_tb}
\dot{P}_{\rm MB} = \dot{P}_{\rm grav} T_{\rm MB} ,  
\end{equation} 
where $T_{\rm MB}=41.6\; (f/k_{0.277})^{-2}m_2^{1/3}m_1^{-4/3}P_{5h}^{2}$ \citep[see equations 2 and 3 in][and references therein]{Burderi_10}. 
The gyration radius $k$ is in units of 0.277, and since the CS has a mass of 0.22~\Msun, we adopt a value $k=0.45$ as reported by \cite{wad_24}. The parameter $f$ can range between 0.78 and 1.73 because it is model dependent \citep[see discussion in][]{Burderi_10}, and we estimate $T_{\rm MB}$ for the limit values of $f$ finding that $T_{\rm MB}$ ranges between 5.8 and 29. The corresponding orbital period derivative associated with the magnetic braking, $\dot{P}_{\rm MB}$, is therefore between 
$-8.2 \times 10^{-13}$~s/s and $-1.6 \times 10^{-13}$~s/s. 
The contribution of the angular momentum loss via GR and MB to the orbital period derivative yields $\dot{P}$ between $-8.5  \times 10^{-13}$~s/s and $-1.9 \times 10^{-13}$~s/s, still compatible with what is obtained from our analysis adopting $T_{\rm MB}=5.8$.

\subsection{The hierarchical triple system scenario}

From the delays fitting, we obtain that the statistically most significant solution involves a sinusoidal modulation with an eccentricity $e$ of  $0.38\pm0.17$ and a modulation period of $17.1 \pm 1.5$~years. The delay modulation could be caused by a third body orbiting around  \xsource, where the modulation period corresponds to the revolution period of the third body around the binary system.
 We can estimate the mass of the third body under the assumption that its orbit is co-planar with that of the binary system. We estimate the separation between the center of mass (CM) of the binary system and that of the triple hierarchical system to be $a_x \sin i = A c$, where $c$ represents the speed of light, and $A$ is the amplitude of the modulation;  $a_x \sin i = 1.8 \times 10^{11}$~cm in our case. The mass $M_3$ of the third body is given by 
\begin{equation}
\label{eq:orbit}
\frac{M_3 \sin i}{(M_3+M_{bin})^{2/3}}=\left(\frac{4\pi^2}{G}\right)^{1/3}\frac{a_x\sin i}{P_m^{2/3}},
\end{equation}  
where $M_{bin}$ is the mass of the binary system, $G$ the gravitational constant and $P_m$ the periodic modulation inferred from the delays fit. 
For an inclination angle of 78\degree (we assume that the third body orbit lies on the binary system orbital plane), $M_1=1.4$~\Msun and $M_2=0.22$~\Msun, we infer that $M_3\simeq 2.7$~M$_J$, where M$_J$ indicates the Jupiter mass. 
The separation between the third body and the CM of the hierarchical triple system is $a_3 = a_x \sin i \; M_{bin}/M_3\simeq 1.1 \times 10^{14}$~cm, corresponding to 7.5~AU, approximately equivalent to a distance similar to that between the Sun and halfway between Jupiter and Saturn.
Finally, at the periastron passage, the separation between the CM and the third body  is $d=a_x \sin i \;(1-e) \; M_{bin}/M_3=7.1 \times 10^{13}$~cm, a distance significantly exceeding the orbital separation of the binary system, represented by $a=9.1 \times 10^{10}$~cm.

The existence of third celestial bodies in orbit around binary systems, particularly those containing a compact object, is being increasingly confirmed by recent discoveries. Notable among these findings is the proposed detection of a third body with a mass of approximately 45 M$_J$ (0.043 M${\odot}$) by \cite{Iaria_2015} and \cite{Iaria_21} orbiting  the dipping source XB 1916-053. Concurrently, \cite{iaria_18} have revealed a 22 M$_J$ mass third body orbiting the eclipsing source MXB 1659-298. The precedent for such detections was arguably set by \cite{Sigurdsson_93}, who proposed the existence of a sub-Jovian mass planet in orbit around the binary system of the millisecond radio pulsar PSR 1620-26 within the globular cluster M4. Subsequent \textit{Hubble Space Telescope} observations allowed to refine this model, positing the third body as a planet with a mass of $2.5 \pm 1.0$ M$_J$ in orbit around a binary system of a millisecond pulsar and a white dwarf companion \citep{Sigurdsson_03}.

\cite{Bailes_11} discovered  a Jupiter-sized chtonian body   composed primarily of carbon and oxygen in a close orbit around the millisecond pulsar PSR J1719-1438, at a  distance of 0.004 AU. 
Such observations are in line with the seminal discovery of planets orbiting PSR 1257+12 \cite{Wolszczan_92}, which has spawned a multitude of hypotheses regarding planetary formation in the vicinity of pulsars.

\cite{Pod_93}  discussed that planetary bodies associated with neutron stars are supposed to form during one of three distinct  periods: i) within a protoplanetary disc during the initial star formation (first-generation); ii) from a fallback disc composed of debris of the supernova explosion (second-generation); iii) or within an accretion disc that forms from matter transferred to the neutron star from the companion star (third-generation). Planets from the first-generation, if they exist, are presumed to be either destroyed  or to have their trajectories severely altered in the wake of a supernova event. The existence of planets orbiting millisecond pulsars challenges the hypothesis of second-generation formation, as the spin acceleration implies a proximal companion star, within an orbital distance of roughly 1 AU, and this would likely perturb  planets orbits. This is corroborated by the instance of PSR B1257+12, which is a fully recycled pulsar, implying its companion  is a low-mass star and had experienced Roche lobe overflow during its main-sequence life stage, necessitating an orbital proximity not exceeding 1 AU.

Therefore, the prevailing hypothesis for the formation of planets orbiting isolated millisecond pulsars supports the idea of third-generation formation, where planets form from the residual matter of the companion star.
 However, this  evolutionary channel  clashes with the findings of this study, where we have an X-ray binary system undergoing mass transfer from a main-sequence
 companion star. Similarly, this evolutionary path cannot easily explain the three-body system of the millisecond pulsar PSR 1620-26, where the main companion star is a white dwarf. Therefore, one possibility is that these planets are solitary planets that were gravitationally captured by the binary system \citep{Podsiadlowski_91}, which is very unlikely for Galactic field systems.
 Alternatively, it is possible that these planets emerge from a circumbinary disk, formed from material expelled by the companion star during a period of non-conservative mass transfer of the LMXB 
\citep{Tavani_92}.

\subsection{Limitations of the Applegate mechanism in explaining orbital period modulation in \xsource}

 Given that the CS is a low-mass star (0.22~\Msun), we expect its magnetic activity to be intense. However, \xsource  exhibits persistent emission with mass transfer via  Roche lobe overflow, so the luminosity from the outer regions of the accretion disk may overshadow that of the CS in the visible band, hindering a detailed study of the latter \cite{Ratti_10}. Consequently, the magnetic properties of the CS in \xsource are currently unknown.

Nevertheless, an alternative explanation of the modulation of approximately 17 years observed in the orbital delays could be given by a gravitational coupling of the orbit with variations in the shape of the magnetically active CS.
These variations are believed to result from the torque exerted by the magnetic activity associated with a subsurface magnetic field in the CS, interacting with its convective envelope. The convective envelope initiates a periodic exchange of angular momentum between the inner and outer regions of the CS, leading to alterations in its gravitational quadrupole moment \citep{Applegate_92,Applegate_94}.

In this instance, the deduced periodicity of 6249 days and amplitude of 6.1~s translates to an orbital period variation of $\Delta P/P\simeq 2.2\times 10^{-5}$. Under the reasonable assumption that the CS fills its Roche lobe, and consequently, its radius coincides with the Roche-lobe radius, we can calculate the angular momentum transfer required to induce the observed orbital period change to be approximately $\Delta J \simeq 4.8 \times 10^{44}$~g cm$^2$/s. 

The non-synchronicity of the companion, expressed by the ratio $\Delta \Omega/\Omega$, is $6.7\times 10^{-5}$, where $\Omega$ is the orbital angular velocity of the binary system and 
$\Delta \Omega$ represents the variation in the orbital angular velocity required to induce the change in the orbital period $\Delta P$. The value of $\Delta \Omega/\Omega$ is obtained assuming that a thin shell of mass $M_s$ is  10\% of the CS mass, where the shell represents the limited mass in the outer part of the star that governs the quadrupole moment. For $M_s$ mass larger than 10\%, the angular momentum transfer mechanism ceases to function \citep{Applegate_92}. 

The variable part of the CS luminosity  necessary to fuel the variations in the gravitational quadrupole is $\Delta L=2.9 \times 10^{-5}$~L$_{\odot}$. To estimate the required fraction of the CS luminosity to achieve the observed modulation in the delays, we deduce the CS luminosity using the relationship $L/L_{\odot}= 0.231\; (M_2/\Msun)^{2.61}$, applicable to low-mass stars. We find that $L\simeq 4.4 \times 10^{-3}$~L$_{\odot}$ and $\Delta L \simeq 0.007$. 
To achieve these brightness variations, we would need a shell mass of 0.7\% of the CS mass, while the available budget should be around 10\% of the CS's luminosity. This result appears odd with the model, so we conclude that this is unlikely to be the correct explanation.

\section{Conclusions}
We used the mid-eclipse times reported by \cite{Chetana_11} and \cite{chetana_22} to offer a different interpretation of the orbital residuals of \xsource. By fitting the delays, we find a periodic modulation, close to 17 years, described by an eccentric sinusoidal modulation with an eccentricity of about 0.38 and an amplitude of 6.1~s.  

The likely scenario describing the sinusoidal modulation involves a third body orbiting around the binary system with a revolution period of 17 years and an orbital eccentricity of 0.38. From these parameters, we deduce that the mass of the third body is 2.7~M$_J$. The presence of a third body around a binary system has also been discussed for the eclipsing LMXB  MXB 1659-298 \citep{iaria_18}; in that case, the third body has a mass exceeding 21~M$_J$. Moreover, the presence of a third body has also been considered by \cite{Iaria_2015} for the ultra-compact LMXB  XB 1916-053; in this latter case, the mass of the third body exceeds 45~M$_J$.   

It is important to emphasize that the presence of a third orbiting body does not contract or expand the orbital period of a binary system. Indeed, the motion of the third body influences the position of the center of mass of the binary system, thereby affecting the arrival times of eclipses. The derivative of the orbital period obtained in this work is not affected at zero order by the presence of the third body;  it reflects the expected evolution of an LMXB system given by the mass transfer and the loss of angular momentum via magnetic braking and gravitational radiation in the case of a conservative mass transfer.

Our interpretation of the results, statistically equivalent to that obtained by \cite{chetana_22}, rules out the presence of discontinuities in the derivative of the orbital period, for which, to date, there is no theoretical model for their interpretation. Instead, our result aligns with the evolution of a binary system in which the CS of 0.22~\Msun is in thermal equilibrium, and conservative mass transfer tends to contract the binary system.

\begin{acknowledgements}  The authors acknowledge financial support  from  PRIN-INAF 2019 with the project "Probing the geometry of accretion: from theory to observations" (PI: Belloni). WL conducted this research during, and with the support of, the Italian national inter-university PhD program in Space Science and Technology.

\end{acknowledgements}
\bibliographystyle{aa}
\bibliography{biblio}

\begin{thebibliography}{39}
\expandafter\ifx\csname natexlab\endcsname\relax\def\natexlab#1{#1}\fi

\bibitem[{{Applegate}(1992)}]{Applegate_92}
{Applegate}, J.~H. 1992, \apj, 385, 621

\bibitem[{{Applegate} \& {Shaham}(1994)}]{Applegate_94}
{Applegate}, J.~H. \& {Shaham}, J. 1994, \apj, 436, 312

\bibitem[{{Bailes} {et~al.}(2011){Bailes}, {Bates}, {Bhalerao}, {Bhat}, {Burgay}, {Burke-Spolaor}, {D'Amico}, {Johnston}, {Keith}, {Kramer}, {Kulkarni}, {Levin}, {Lyne}, {Milia}, {Possenti}, {Spitler}, {Stappers}, \& {van Straten}}]{Bailes_11}
{Bailes}, M., {Bates}, S.~D., {Bhalerao}, V., {et~al.} 2011, Science, 333, 1717

\bibitem[{{Brookshaw} \& {Tavani}(1993)}]{Brookshow_93}
{Brookshaw}, L. \& {Tavani}, M. 1993, \apj, 410, 719

\bibitem[{{Burderi} {et~al.}(2010){Burderi}, {Di Salvo}, {Riggio}, {Papitto}, {Iaria}, {D'A{\`\i}}, \& {Menna}}]{Burderi_10}
{Burderi}, L., {Di Salvo}, T., {Riggio}, A., {et~al.} 2010, \aap, 515, A44

\bibitem[{{Chen}(2017)}]{Chen_17}
{Chen}, W.-C. 2017, \mnras, 464, 4673

\bibitem[{{di Salvo} {et~al.}(2008){di Salvo}, {Burderi}, {Riggio}, {Papitto}, \& {Menna}}]{disalvo_08}
{di Salvo}, T., {Burderi}, L., {Riggio}, A., {Papitto}, A., \& {Menna}, M.~T. 2008, \mnras, 389, 1851

\bibitem[{{Eggleton}(1983)}]{Eggleton_83}
{Eggleton}, P.~P. 1983, \apj, 268, 368

\bibitem[{{Frank} {et~al.}(1987){Frank}, {King}, \& {Lasota}}]{Frank_87}
{Frank}, J., {King}, A.~R., \& {Lasota}, J.~P. 1987, \aap, 178, 137

\bibitem[{{Galloway} {et~al.}(2008){Galloway}, {Muno}, {Hartman}, {Psaltis}, \& {Chakrabarty}}]{Galloway_08}
{Galloway}, D.~K., {Muno}, M.~P., {Hartman}, J.~M., {Psaltis}, D., \& {Chakrabarty}, D. 2008, \apjs, 179, 360

\bibitem[{{Iaria} {et~al.}(2015){Iaria}, {Di Salvo}, {Gambino}, {Del Santo}, {Romano}, {Matranga}, {Galiano}, {Scarano}, {Riggio}, {Sanna}, {Pintore}, \& {Burderi}}]{Iaria_2015}
{Iaria}, R., {Di Salvo}, T., {Gambino}, A.~F., {et~al.} 2015, \aap, 582, A32

\bibitem[{{Iaria} {et~al.}(2018){Iaria}, {Gambino}, {Di Salvo}, {Burderi}, {Matranga}, {Riggio}, {Sanna}, {Scarano}, \& {D'A{\`\i}}}]{iaria_18}
{Iaria}, R., {Gambino}, A.~F., {Di Salvo}, T., {et~al.} 2018, \mnras, 473, 3490

\bibitem[{{Iaria} {et~al.}(2021){Iaria}, {Sanna}, {Di Salvo}, {Gambino}, {Mazzola}, {Riggio}, {Marino}, \& {Burderi}}]{Iaria_21}
{Iaria}, R., {Sanna}, A., {Di Salvo}, T., {et~al.} 2021, \aap, 646, A120

\bibitem[{{Jain} \& {Paul}(2011)}]{Chetana_11}
{Jain}, C. \& {Paul}, B. 2011, \mnras, 413, 2

\bibitem[{{Jain} {et~al.}(2022){Jain}, {Sharma}, \& {Paul}}]{chetana_22}
{Jain}, C., {Sharma}, R., \& {Paul}, B. 2022, \mnras, 517, 2131

\bibitem[{{Knigge} {et~al.}(2011){Knigge}, {Baraffe}, \& {Patterson}}]{Knigge_11}
{Knigge}, C., {Baraffe}, I., \& {Patterson}, J. 2011, \apjs, 194, 28

\bibitem[{{Markwardt} {et~al.}(1998){Markwardt}, {Marshall}, {Swank}, \& {Takeshima}}]{Mark_98}
{Markwardt}, C.~B., {Marshall}, F.~E., {Swank}, J., \& {Takeshima}, T. 1998, \iaucirc, 6998, 2

\bibitem[{{Markwardt} {et~al.}(2001){Markwardt}, {Swank}, \& {Strohmayer}}]{Mark_01}
{Markwardt}, C.~B., {Swank}, J.~H., \& {Strohmayer}, T.~E. 2001, in American Astronomical Society Meeting Abstracts, Vol. 199, American Astronomical Society Meeting Abstracts, 27.04

\bibitem[{{Neece}(1984)}]{Neece_84}
{Neece}, G.~D. 1984, \apj, 277, 738

\bibitem[{{Paczy{\'n}ski}(1971)}]{Pac_71}
{Paczy{\'n}ski}, B. 1971, \araa, 9, 183

\bibitem[{{Podsiadlowski}(1993)}]{Pod_93}
{Podsiadlowski}, P. 1993, in Astronomical Society of the Pacific Conference Series, Vol.~36, Planets Around Pulsars, ed. J.~A. {Phillips}, S.~E. {Thorsett}, \& S.~R. {Kulkarni}, 149--165

\bibitem[{{Podsiadlowski} {et~al.}(1991){Podsiadlowski}, {Pringle}, \& {Rees}}]{Podsiadlowski_91}
{Podsiadlowski}, P., {Pringle}, J.~E., \& {Rees}, M.~J. 1991, \nat, 352, 783

\bibitem[{{Ponti} {et~al.}(2017){Ponti}, {De}, {Mu{\~n}oz-Darias}, {Stella}, \& {Nandra}}]{Ponti_17}
{Ponti}, G., {De}, K., {Mu{\~n}oz-Darias}, T., {Stella}, L., \& {Nandra}, K. 2017, \mnras, 464, 840

\bibitem[{{Ponti} {et~al.}(2012){Ponti}, {Fender}, {Begelman}, {Dunn}, {Neilsen}, \& {Coriat}}]{Ponti_2012}
{Ponti}, G., {Fender}, R.~P., {Begelman}, M.~C., {et~al.} 2012, \mnras, 422, L11

\bibitem[{{Raman} {et~al.}(2018){Raman}, {Maitra}, \& {Paul}}]{Raman_18}
{Raman}, G., {Maitra}, C., \& {Paul}, B. 2018, \mnras, 477, 5358

\bibitem[{{Rappaport} {et~al.}(1983){Rappaport}, {Verbunt}, \& {Joss}}]{rappaport_83}
{Rappaport}, S., {Verbunt}, F., \& {Joss}, P.~C. 1983, \apj, 275, 713

\bibitem[{{Ratti} {et~al.}(2010){Ratti}, {Bassa}, {Torres}, {Kuiper}, {Miller-Jones}, \& {Jonker}}]{Ratti_10}
{Ratti}, E.~M., {Bassa}, C.~G., {Torres}, M.~A.~P., {et~al.} 2010, \mnras, 408, 1866

\bibitem[{{Ruderman} {et~al.}(1989){Ruderman}, {Shaham}, {Tavani}, \& {Eichler}}]{ruderman_89}
{Ruderman}, M., {Shaham}, J., {Tavani}, M., \& {Eichler}, D. 1989, \apj, 343, 292

\bibitem[{{Sigurdsson}(1993)}]{Sigurdsson_93}
{Sigurdsson}, S. 1993, \apjl, 415, L43

\bibitem[{{Sigurdsson} {et~al.}(2003){Sigurdsson}, {Richer}, {Hansen}, {Stairs}, \& {Thorsett}}]{Sigurdsson_03}
{Sigurdsson}, S., {Richer}, H.~B., {Hansen}, B.~M., {Stairs}, I.~H., \& {Thorsett}, S.~E. 2003, Science, 301, 193

\bibitem[{{Tailo} {et~al.}(2018){Tailo}, {D'Antona}, {Burderi}, {Ventura}, {di Salvo}, {Sanna}, {Papitto}, {Riggio}, \& {Maselli}}]{Tailo_18}
{Tailo}, M., {D'Antona}, F., {Burderi}, L., {et~al.} 2018, \mnras, 479, 817

\bibitem[{{Tavani} \& {Brookshaw}(1992)}]{Tavani_92}
{Tavani}, M. \& {Brookshaw}, L. 1992, \nat, 356, 320

\bibitem[{{van den Heuvel}(1994)}]{heuvel_94}
{van den Heuvel}, E.~P.~J. 1994, in Saas-Fee Advanced Course 22: Interacting Binaries, 263--474

\bibitem[{{Verbunt}(1993)}]{verbunt_93}
{Verbunt}, F. 1993, \araa, 31, 93

\bibitem[{{Wadhwa} {et~al.}(2024){Wadhwa}, {Landin}, {Kosti{\'c}}, {Vince}, {Arbutina}, {De Horta}, {Filipovi{\'c}}, {Tothill}, {Petrovi{\'c}}, \& {Djura{\v{s}}evi{\'c}}}]{wad_24}
{Wadhwa}, S.~S., {Landin}, N.~R., {Kosti{\'c}}, P., {et~al.} 2024, \mnras, 527, 1

\bibitem[{{Wolff} {et~al.}(2002){Wolff}, {Hertz}, {Wood}, {Ray}, \& {Bandyopadhyay}}]{Wolff_02}
{Wolff}, M.~T., {Hertz}, P., {Wood}, K.~S., {Ray}, P.~S., \& {Bandyopadhyay}, R.~M. 2002, \apj, 575, 384

\bibitem[{{Wolff} {et~al.}(2007){Wolff}, {Wood}, \& {Ray}}]{Wolff_07}
{Wolff}, M.~T., {Wood}, K.~S., \& {Ray}, P.~S. 2007, \apjl, 668, L151

\bibitem[{{Wolszczan} \& {Frail}(1992)}]{Wolszczan_92}
{Wolszczan}, A. \& {Frail}, D.~A. 1992, \nat, 355, 145

\bibitem[{{Younes} {et~al.}(2009){Younes}, {Boirin}, \& {Sabra}}]{younes_09}
{Younes}, G., {Boirin}, L., \& {Sabra}, B. 2009, \aap, 502, 905

\end{thebibliography}
\end{document}